\documentclass[aps,prb,twocolumn,superscriptaddress,showpacs]{revtex4}
\usepackage{graphicx}

\begin{document}


\title{In-plane uniaxial anisotropy rotations in (Ga,Mn)As thin films}


\
\author{M. Sawicki}
\email{mikes@ifpan.edu.pl} \affiliation{Institute of Physics,
Polish Academy of Sciences, al.~Lotnik\'ow 32/46, PL-02668
Warszawa, Poland}
\author{K-Y. Wang}
 \affiliation{School of Physics and Astronomy, University of
Nottingham, Nottingham NG7 2RD, UK}
\author{K.W. Edmonds}
\affiliation{School of Physics and Astronomy, University of
Nottingham, Nottingham NG7 2RD, UK}
\author{R.P. Campion}
\affiliation{School of Physics and Astronomy, University of
Nottingham, Nottingham NG7 2RD, UK}
\author{C.R. Staddon}
\affiliation{School of Physics and Astronomy, University of
Nottingham, Nottingham NG7 2RD, UK}
\author{N.R.S. Farley}
\affiliation{School of Physics and Astronomy, University of
Nottingham, Nottingham NG7 2RD, UK}
\author{\\C.T. Foxon}
\affiliation{School of Physics and Astronomy, University of
Nottingham, Nottingham NG7 2RD, UK}
\author{E. Papis}
\affiliation{Institute of Electon Technology, al.~Lotnik\'ow
32/46, 02-668 Warszawa, Poland}
\author{E. Kami\'nska}
\affiliation{Institute of Electon Technology, al.~Lotnik\'ow
32/46, 02-668 Warszawa, Poland}
\author{A. Piotrowska}
\affiliation{Institute of Electon Technology, al.~Lotnik\'ow
32/46, 02-668 Warszawa, Poland}
\author{T. Dietl}
\affiliation{Institute of Physics, Polish Academy of Sciences,
al.~Lotnik\'ow 32/46, PL-02668 Warszawa, Poland}
\affiliation{ERATO Semiconductor Spintronics Project, al.
Lotnik\'ow 32/46,~PL-02668~Warszawa, Poland\\ and Institute of
Theoretical Physics, Warsaw University, PL-00681 Warszawa, Poland}
\author{B.L. Gallagher}
\affiliation{School of Physics and Astronomy, University of
Nottingham, Nottingham NG7 2RD, UK}

\date{\today}

\begin{abstract}

We show, by SQUID magnetometry, that in (Ga,Mn)As films the
in-plane uniaxial magnetic easy axis is consistently associated
with particular crystallographic directions and that
it~can~be~rotated from the $[\bar{1}10]$ direction to the $[110]$
direction by low temperature annealing. We show that this behavior
is hole--density--dependent and does not originate from surface
anisotropy. The presence of uniaxial anisotropy as well its
dependence on the hole-concentration and temperature can be
explained in terms of the p-d Zener model of the ferromagnetism
assuming a small trigonal  distortion.

\end{abstract}

\pacs{75.50.Pp, 75.30.Gw, 75.70.-i}

\maketitle


Carrier-mediated ferromagnetism in the magnetic semiconductor
(Ga,Mn)As has attracted considerable interest due to its potential
application in spintronics, where the electron spin is used to
carry information.\cite{Spintronics} It has been known since early
works that (Ga,Mn)As films show rather strong magnetic anisotropy,
which is largely controlled by epitaxial strain. The magnetic easy
axis orients out-of-plane and in-plane under tensile and
compressive biaxial strains,
respectively.\cite{Mune93,Shen97,Liu03} This is well understood
within the framework of the p-d Zener model of the hole-mediated
ferromagnetism.\cite{Diet00,Diet01a,Abol01,Sawi02} More recent
works have shown, that these systems also exhibit a strong {\em
in-plane} uniaxial magnetic
anisotropy.\cite{Sawi02,Tang03,Liu03,Hrab02,Welp03} This shows the
existence of a symmetry breaking mechanism, whose microscopic
origin has not yet been identified. Since the magnetic anisotropy
will have a marked influence on spin injection and
magnetotunnelling devices,\cite{Tana01} it is important to develop
a greater understanding of this property, and the methods for its
control.

Here we study the in-plane magnetic anisotropy in a series of
(Ga,Mn)As thin films, and show that the easy direction is either
$[110]$, $[100]$, or $[\bar{1}10]$, depending on the carrier
density and also on the temperature. We demonstrate, in
particular, that the orientation of the uniaxial axis changes by
$90^{\circ}$ after annealing, if the hole concentration is
increased above $\sim6\times$10$^{20}$ cm$^{-3}$.  This finding
suggests the in-plane magnetization direction can be controlled by
changing the hole concentration, e.g. by gating or illumination.
Furthermore, studies of magnetic response after subsequent etching
steps shows that the uniaxial anisotropy is {\em not} a surface
effect. By extending the previous theory\cite{Diet01a} to the case
of arbitrary deformation, we find that the observed in-plane
uniaxial anisotropy can be accounted for by a trigonal distortion
$\varepsilon_{xy} \approx 0.05$\%.


A wide range of 50~nm thick Ga$_{1-x}$Mn$_x$As thin films with
$1.7 \leq x \leq 9$\%, and of different thickness ($10 \leq d \leq
100$ nm) for fixed Mn concentration $x = 6.7$\% were grown on
GaAs(001) substrates by low temperature (180--300$^{\circ}$C)
molecular beam epitaxy using As$_2$.\cite{Camp03} The Mn
concentrations are determined from the Mn/Ga flux ratio,
calibrated by secondary ion mass spectroscopy (SIMS) \ on 1~$\mu$m
thick samples grown under otherwise identical conditions. The
crystallographic orientation of the wafer is determined from RHEED
measurements during growth, with measurements along the $[110]$
direction giving the $2\times$~pattern. For the magnetometry
studies the material is cleft into, typically, $4 \times 5$~mm$^2$
rectangles, whose precise crystallographic orientation is
reconfirmed by Laue back-reflection x-ray diffraction, which can
unambiguously distinguish between diagonal $[110]$ and
$[\bar{1}10]$ directions in GaAs.\cite{Fews00} For electrical
investigations, Hall bar structures are defined lithographically.


The hole concentration $p$ is obtained from magnetoresistance and
Hall measurements in magnetic fields up to 16~T and temperatures
down to 0.3~K, using the routine described in detail
previously.\cite{Edmo02a,Wang04} All samples show metallic
conductivity, so that the hole concentrations can be obtained
reasonably accurately by this method. The magnetic anisotropy is
assessed using a custom built low-field SQUID magnetometer.
Detailed information about magnetic anisotropy is obtained both
from {\it m(H)} curves recorded for various crystallographic
orientations,\cite{Wang04x} and from the temperature dependence of
the remnant magnetization $M_{\mbox{\tiny{REM}}}$. For the latter
the sample is cooled down through $T_{\mbox{\tiny C}}$ in an
external magnetic field of 1000~Oe, which is at least few times
above any coercive field in the studied material. Then the field
is removed at $T = 5$~K and the $M_{\mbox{\tiny{REM}}}$ component
along the field-cooled direction is measured as a function of
increasing temperature. The extremely low value of the external
field trapped in our superconducting coil (typical trapped field
stays below 0.1~Oe after a trip to 1000~Oe) allows the sample
magnetization to rotate exactly to the nearest easy direction set
only by the torques exerted on $M$ by internal anisotropy
field(s). In general, this procedure, if repeated for the main
crystallographic orientations, allows unambiguous determination of
the orientation of $M$ across the whole temperature range up to
$T_{\mbox{\tiny C}}$. Following the electrical and magnetic
measurements, the samples are annealed at 190$^{\circ}$C in
air\cite{Edmo02b} and then re-measured. Low temperature annealing
is a well--established procedure for promoting outdiffusion of
compensating Mn interstitials from the layers\cite{Edmo04} thus
resulting in an increased hole concentration and Curie
temperature.


Remanent magnetization $M_{\mbox{\tiny{REM}}}(T)$, for two
representative samples with Mn concentration 2.2\% and 5.6\%
respectively, are shown in Figs.~1 and 2. It is well established
that for typical hole densities (Ga,Mn)As/GaAs exhibits an
in-plane magnetic anisotropy which is determined by the
superposition of two components: a biaxial, cubic-like anisotropy
with $\langle 100\rangle$  the easy axes, plus a  uniaxial
anisotropy with $\langle 110\rangle$ the easy
axes.\cite{Sawi02,Tang03,Liu03,Hrab02,Welp03} The former is a
direct consequence of the spin anisotropy of the hole liquid,
originating from a strong spin-orbit coupling in the valence band.
This coupling transfers all the complexities of the valence band
physics into the Mn ions subsystem. As a result the strength of
the cubic anisotropy strongly depends on epitaxial stress, hole
concentration and temperature.\cite{Diet01a,Abol01} The second
in-plane anisotropy component, the uniaxial term, is not expected
from the above model since its presence is precluded by general
symmetry considerations in the biaxially strained zinc blende
structure of (Ga,Mn)As. It has been suggested that these uniaxial
properties may originate from the symmetry lowering
(D$_{\mbox{\tiny{2d}}}$ $\rightarrow$ C$_{\mbox{\tiny{2v}}}$) due
to the lack of top-bottom symmetry in (Ga,Mn)As epilayers,
\cite{Sawi02} or as a consequence of the anisotropic GaAs surface
reconstruction during growth.\cite{Welp03}

\begin{figure}
\includegraphics[width=3.2in]{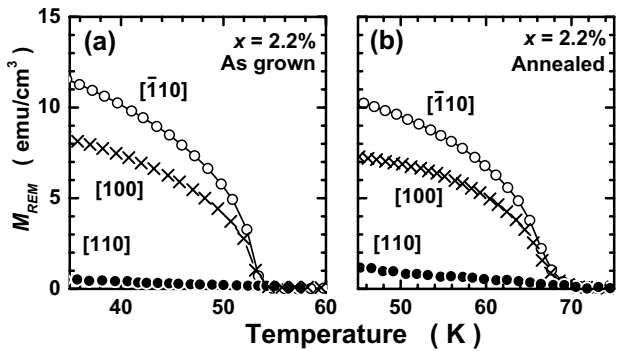}
\caption{\label{fig1} High temperature dependence of the three
major in-plane components of the remnant magnetization in
Ga$_{0.978}$Mn$_{0.022}$As before (\textbf{a}) and
after~annealing~(\textbf{b}).}
\end{figure}
\begin{figure}
\includegraphics[width=2.7in]{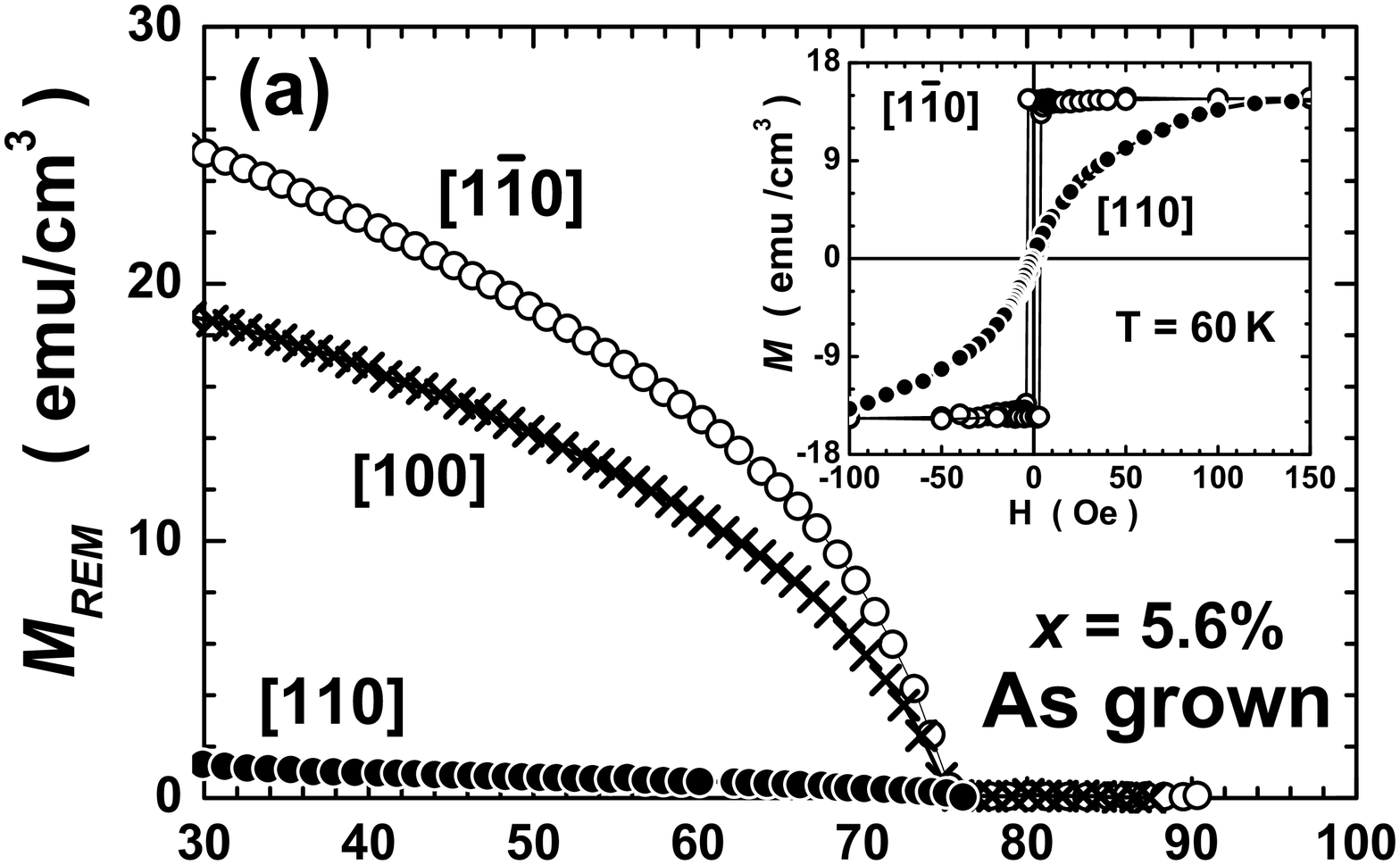}
\includegraphics[width=2.7in]{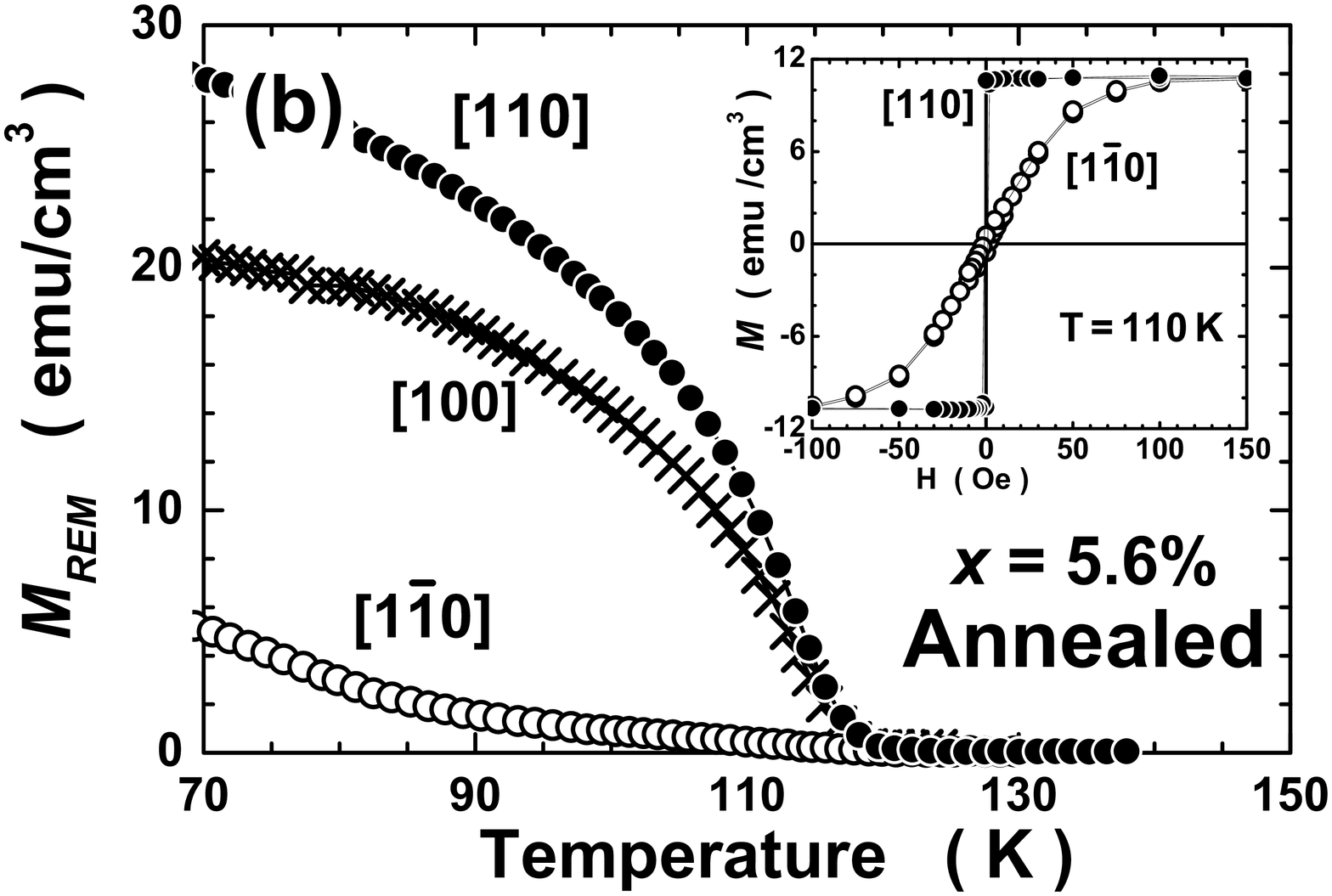}
\caption{\label{fig2}  High temperature dependence of the three
major in-plane components of the remnant magnetization in
Ga$_{0.944}$Mn$_{0.056}$As before (\textbf{a}) and after annealing
(\textbf{b}). Insets: magnetization curves at temperatures close
to $T_{\mbox{\tiny C}}$ measured for $[110]$ and $[\bar{1}10]$
orientations. Note the switch of the magnetic easy axis from
$[\bar{1}10]$ to $[110]$ upon annealing.}
\end{figure}

It is usually found that the cubic term is dominant at low
temperatures. However, on increasing temperature the strength of
the uniaxial term decreases much more slowly than that of the
cubic one (the uniaxial anisotropy field, $H_U$, decreases as $M$
while the cubic one, $H_C$, falls as $M^{3}$), and at elevated
temperatures the uniaxial term becomes firstly comparable with,
and soon after, stronger than the cubic contribution. In the
latter case the magnetic anisotropy of the layers is solely
determined by the uniaxial term. Note that the temperature
corresponding to the crossover from cubic to uniaxial behaviour
will be determined by the hole density and the precise form of
$M(T)$. This crossover is not necessarily close to
$T_{\mbox{\tiny{C}}}/2$.\cite{Welp03} In the present study we
focus on this high temperature regime, in which
$M_{\mbox{\tiny{REM}}}$ is firmly locked into the easy magnetic
direction, which allows an unambiguous determination of the easy
magnetic axis.

The data of Figs.~1(a) and 2(a) clearly show that the
magnetization is locked into the $[\bar{1}10]$ direction for a
wide temperature range below $T_{\mbox{\tiny C}}$ for the as-grown
samples: along this direction $M_{\mbox{\tiny{REM}}}$ is the
largest, and Brillouin-like; $M_{\mbox{\tiny{REM}}}$ along $[110]$
is (very) small, $M_{\mbox{\tiny{REM}}}$ along $[100]$ is
identical with $[010]$ and both are reduced by approximately
$\cos(45^{\circ})$. This shows that the uniaxial anisotropy is
dominant at these temperatures, as discussed above. A similar
behavior is observed in all as-grown samples studied here.

\begin{figure}
\includegraphics[width=3.0in]{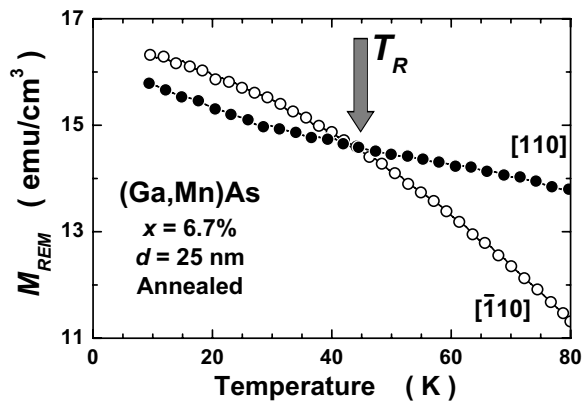}
\caption{\label{fig3} Temperature dependence of the $[\bar{1}10]$
and $[110]$ in-plane components of the remnant magnetization in
Ga$_{0.933}$Mn$_{0.067}$As annealed layer. Note the switch of the
uniaxial easy direction at temperature $T_R \simeq 45$~K.}
\end{figure}
At low Mn concentrations, annealing results in a relatively small
increase in $T_{\mbox{\tiny C}}$ and has no qualitative effect on
the magnetic anisotropy, as shown in Fig.~1(b) for  $x=2.2$\%. A
very different behavior is observed at higher concentrations. As
shown in Fig.~2 for $x=5.6$\%, after annealing the easy axis is
rotated by $90^{\circ}$ to the $[110]$ direction. This rotation of
the uniaxial easy axis is confirmed by $M(H)$ curves in this
temperature range, shown in the insets. Before annealing, the
$M(H)$ curves are square along the $[\bar{1}10]$ direction and
elongated along the $[110]$ direction, and this situation is
reversed after annealing. Interestingly, we also observe the
temperature-induced reorientation of the easy axis as shown in
Fig.~3. In this case, the easy axis points in the $[\bar{1}10]$
direction at low temperatures but assumes the $[110]$ at higher
temperatures.

\begin{figure}
\includegraphics[width=3.0in]{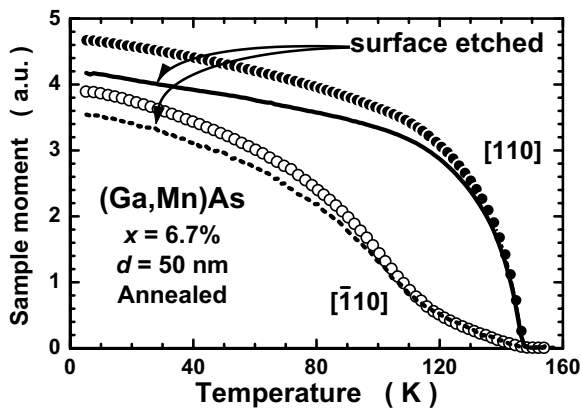}
\caption{\label{fig4}  Temperature dependence of the remnant
magnetic moment in Ga$_{0.933}$Mn$_{0.067}$As annealed layer
measured along uniaxially easy $[110]$ and hard $[\bar{1}10]$
directions before (points) and after (lines) surface etching.}
\end{figure}
Low temperature annealing of (Ga,Mn)As is known to cause an
out-diffusion of Mn interstitials resulting in Mn surface
aggregation, which could in principle affect the magnetic
anisotropy. However, we rule this out as the origin of the
observed reorientation by etching the surface of the layers in
$\mbox{H}_2\mbox{SO}_4 + \mbox{H}_2\mbox{O}_2 +
\mbox{H}_2\mbox{O}$ 1:8:1000. In Fig.~4 we plot
$M_{\mbox{\tiny{REM}}}(T)$ curves for an annealed $x=6.7$\% layer
of initial thickness 50~nm, before and after etching away
approximately 10\% of the layer. Aside from a reduction in the
total magnetic moment of the layer due to the reduced thickness,
no changes in either the magnetic anisotropy or the Curie
temperature are observed. In fact we etched this layer several
more times, removing finally 50\% of the layer, without detecting
any noticeable changes. This rules out any influence of
surface-accumulated Mn, illustrates the high vertical uniformity
in the annealed material, and also points strongly to a mechanism
built into the \emph{bulk} of the layer.

\begin{figure}
\includegraphics[width=3.0in]{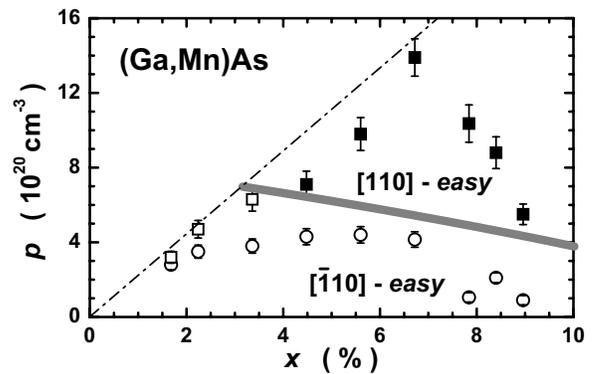}
\caption{\label{fig5} Hole concentration versus Mn concentration
for the series of 50~nm thick samples. Open symbols mark samples
with the uniaxial easy axis oriented along the $[\bar{1}10]$
direction, full symbols denote samples exhibiting the easy axis
along $[110]$. The dashed line marks the compensation free
\emph{p}--type Mn doping level in (Ga,Mn)As. The thick grey line
separates two regions of hole densities where, independently of
being annealed or not, the layers consistently show the same
crystallographic alignment of the uniaxial easy axis.}
\end{figure}
We therefore ascribe the rotation of the easy magnetic axis to the
increase of the hole concentration on annealing. This will modify
the relative occupancies of the valence sub-bands of the GaAs
host, which (at least for the case of cubic anisotropy) make
competing contributions to the magnetic
anisotropy.\cite{Diet01a,Abol01} Figure 5 plots the measured hole
concentration versus the Mn concentration for the series of 50~nm
thick samples. Open symbols mark samples where the easy axis is
oriented along the $[\bar{1}10]$ direction, while for closed
symbols the easy axis is along $[110]$. It can be seen that the
easy axis rotates by $90^{\circ}$ on annealing for samples with
$x\gtrsim 4$\%. By inspection of Fig.~5, we can assign a threshold
value of $p$ of approximately 6$\times$10$^{20}$ cm$^{-3}$, above
which the easy axis orients along $[110]$. This finding, together
with the etching experiment discussed above demonstrate that the
in-plane uniaxial anisotropy depends upon the bulk film
parameters. A recent study showing that the uniaxial anisotropy
field is thickness--independent in the range 0.2~$\mu$m to
6.8~$\mu$m gives further support to this argument.\cite{Welp04}
Thus, this anisotropy is caused by a symmetry lowering mechanism
existing inside the film.

\begin{figure}
\includegraphics[width=3.0in]{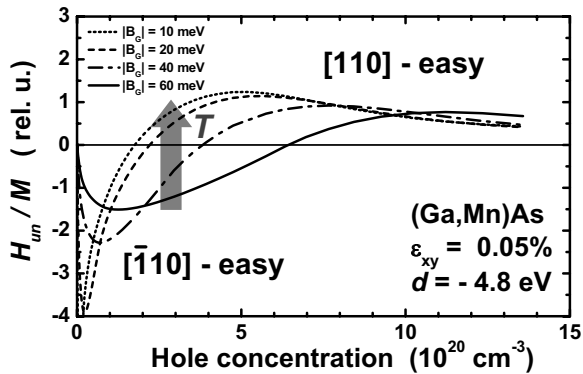}
\caption{\label{fig6} In-plane uniaxial anisotropy field (in SI
units and normalized to sample saturation magnetization) versus
hole density computed for various valence-band spin-splittings in
(Ga,Mn)As. The thick arrow illustrates the possibility of
$[\bar{1}10] \Rightarrow [110]$ easy axis rotation on increasing
temperature as observed experimentally and depicted on Fig.~3.}
\end{figure}
In order to find out the magnitude of the symmetry lowering
perturbation that would explain our findings we incorporate in the
p-d Zener theory of ferromagnetism in tetrahedrally coordinated
semiconductors\cite{Diet00,Diet01a,Abol01} a trigonal distortion
described by the deformation tensor component $\varepsilon_{xy}
\ne 0$. The computed anisotropy field corresponding to the
in-plane uniaxial anisotropy $\mu_oH_{un}$ is shown in Fig.~6 as a
function of the hole concentration and valence-band spin-splitting
parameter $B_G = A_F\beta M(T)/g\mu_B$, where $g=2.0$, $A_F = 1.2$
is the Fermi liquid parameter and $\beta = -0.054$~eVnm$^3$ is the
p-d exchange integral. For these parameters, $|B_G| \approx
30$~meV for $x = 0.05$, which corresponds to the saturation
magnetization $\mu_oM_s = 65$~mT. In the range in question, the
anisotropy field is found to be linear in $\varepsilon_{xy} d$,
where $d= -4.8$~eV is the deformation potential.\cite{Vurg01} In
the relevant region of hole concentration and for
$\varepsilon_{xy} = 0.05$\%, $\mu_oH_{un}$ has the experimental
value of 0.1~T. For a given \emph{x} Fig.~6 shows that a switch in
the easy axis from $[\bar{1}10]$ to $[110]$ will occur on
increasing hole density. As $M$, and so $B_G$, are decreasing
functions of $T$, a uniaxial easy axis reorientation transition
$[\bar{1}10] \Rightarrow [110]$ may occur also on increasing $T$,
see Fig.~6. Therefore the model qualitatively reproduces the
observed change in the easy axis direction as a function of both
hole concentration and temperature.


In summary, we have shown that the easy magnetization axis of
(Ga,Mn)As films can rotate from the $[\bar{1}10]$ direction to the
$[110]$ direction on annealing. We demonstrate that the
orientation of the in-plane uniaxial anisotropy in (Ga,Mn)As is
dependent on the hole concentration. Since the hole concentration
in III-V magnetic semiconductors can be modified by gating or
illumination, this suggests the possibility of a $90^{\circ}$
rotation of the magnetization without application of an external
magnetic field, which could have important applications in
spintronics. Our results show that this behavior does not
originate from surface anisotropy. We demonstrate that the
magnitude of uniaxial anisotropy as well its dependence on the
hole-concentration and temperature can be explained in terms of
the p-d Zener model of the ferromagnetism assuming a small
trigonal--like distortion.  Such a distortion may be associated
with magnetostriction, or may result from a non-isotropic Mn
distribution, caused for instance by the presence of surface
dimers oriented along $[\bar{1}10]$ direction during the epitaxy.
However, the dominating microscopic mechanism that breaks
D$_{\mbox{\tiny{2d}}}$ symmetry of (Ga,Mn)As epitaxial films is to
be elucidated.

Support of EPSRC, EU FENIKS (G5RD-CT-2001-0535), and Polish
PBZ/KBN/044/PO3/2001 projects is gratefully acknowledged.


\end{document}